E. RÓWIŃSKI*#, M. PIETRUSZKA**

# MODIFIED BOHM'S THEORY FOR ABSTRUSE MEASUREMENTS: APPLICATION TO LAYER DEPTH PROFILING USING AUGER SPECTROSCOPY

Modified Bohm's formalism was applied to solve the problem of abstruse layer depth profiles measured by the Auger electron spectroscopy technique in real physical systems. The desorbed carbon/passive layer on an NiTi substrate and the adsorbed oxygen/surface of an NiTi alloy were studied. It was shown that the abstruse layer profiles can be converted to real layer structures using the modified Bohm's theory, where the quantum potential is due to the Auger electron effect. It is also pointed out that the stationary probability density predicts the multilayer structures of the abstruse depth profiles that are caused by the carbon desorption and oxygen adsorption processes. The criterion for a kind of break or "cut" between the physical and unphysical multilayer systems was found. We conclude with the statement that the physics can also be characterised by the abstruse measurement and modified Bohm's formalism.

## 1. Introduction

Auger electron spectroscopy (AES) is now a widely used technique that is used to investigate multilayer systems [1,2]. The Auger process is understood as the relaxation of an atom with a hole in the inner electron shell by the emission of an electron, the Auger electron [2]. The combination of the AES signal and Ar$^+$-ion beam sputter system is called a sputter depth profile [1,3-9]. From the theoretical point of view, such multilayer systems have not been explained correctly. From the experimental point of view, the sputter depth profile can be given by a partial description of a multilayer system (a film deposition). This means that the layer structure of a thin film can usually be observed by the depth profile in an experiment, which is called an abstruse measurement. Therefore, the layer position can be considered to be a natural choice for the selection and numbering. Another example of an abstruse monolayer – a coverage profile is based on the concept of adsorbed ions on sites on the surface of a solid so that the coverage varies from zero to one monolayer given in the adsorption process [1,2,10-12]. However, it was found that such abstruse physical systems can be described by Bohm's theory [13].

In Bohm's theory, a system of quantum particles is described in part by its wave function. This description is completed by a specification of the actual positions of the particles. Bohm re-interpreted the mathematics of Quantum Mechanics and extracted a part of the equation, which he called the quantum potential. The theory is strictly deterministic. In what follows, we do hope that the modified Bohm's theory will be experimentally verified. Therefore, the main question is: How is the abstruse layer profile converted to the real layer structures during the adsorption and desorption processes. To answer this question, we start our discussion with the Bohm's equations [13]. Firstly, we will briefly review the quantum potential due to the Auger electron effect and the Laplace transform formalism. Then, the solution of the transformed problem, with the help of the Laplace transform, can be obtained for the rescaled depth profiles from the experiment.

## 2. Bohm's theory

We start our considerations using Bohm's equations, which can be expressed as

$$\frac{\partial S_i}{\partial t} + V_i + Q_i + \frac{1}{2m_i}[\nabla S_i]^2 = 0 \qquad (1)$$

$$\frac{\partial R_i^2}{\partial t} + div(R_i^2 \nabla S_i / m_i) = 0 \qquad (2)$$

where $Q_i$ is the quantum potential, $S_i$ is the Hamiltonian-Jacobi function, $P_i \sim R_i^2 \sim |\psi|^2$ is the probability density and $\psi$ is the wave function; index $i$ denotes $i^{th}$ particle [13]. If $Q_i$ can be neglected, then the equation reduces to the classical

* UNIVERSITY OF SILESIA, INSTITUTE OF MATERIALS SCIENCE, 75 PUŁKU PIECHOTY 1A, 41-500 CHORZÓW, POLAND
** UNIVERSITY OF SILESIA, FACULTY OF BIOLOGY AND ENVIRONMENT PROTECTION, JAGIELLOŃSKA 32, 40-032 KATOWICE, POLAND
# Corresponding author: edward.rowinski@us.edu.pl

Hamilton-Jacobi equation. This equation describes the trajectories of a particle with the momentum $p_i = m_i v_i = \nabla S_i$.

Without losing generality, we will consider the atom of the $i^{th}$-layer that contains an electron with the coordinate $(z')$, which has an initial state wave function $\psi_{initial}(z')\exp(-iE_o t/\hbar)$ that corresponds to the stationary state with energy $E_o$. The atom is excited by a primary electron bombardment and leads to a new state with the wave function $\psi_f(z')\exp(-iE_f^{**}t/\hbar)$ and the energy $E_f^{**}$, where the symbol "**" denotes the two electron ionisation. In order to make this possible, we assume an additional particle with the coordinate, $\tau_p$ where $p$ denotes the element that will take up the kinetic energy of the outgoing electron $E_o - E_f^{**}$, which is released in the Auger transition. In a simple case, the so-called Bohm's wave function can be expressed by

$$\psi(z', \tau_p, t) = \psi_{initial}(z')\exp(-iE_o t/\hbar)\phi_{initial}(\tau_p, t)$$
$$+ \int_0^t \alpha(t', t)\psi_f(z')\exp(-iE_f t/\hbar)\phi_f(\tau_p, t-t')dt'$$

where $\phi_{initial}(\tau_p, t)$ is the initial function of the Auger electron and $\alpha(t', t)$ can be calculated using the time-dependent perturbation theory [13]. The integrand produces a contribution to the wave function at a small interval of time $dt'$. This corresponds to the source position of the Auger electron $(\tau_p)$, which during the time interval $t - t'$, moves away from the atom very rapidly. For example, we can deal with Auger experiments that produce holes in the valence states. For the core-core valence (or $CC'V$) transition, the Auger-emission cross section will be given by

$$\frac{d\sigma^{CC'V}}{dE} \propto \mathrm{Re}\,\pi^{-1}\int_0^\infty <d_{\tau_p\sigma}(t)d_{\tau_p\sigma}^+(0)> e^{-i(E_c - E_{c'} - E)t}dt$$

where $d_{\tau_p\sigma}$ creates a hole of spin $\sigma$ at the site $\tau_p$; $E_c, E_{c'}$ are the binding energies of the core hole, $E$ is the kinetic energy of the outgoing Auger electron and $<d_{\tau_p\sigma}(t)d_{\tau_p\sigma}^+(0)>$ is the autocorrelation function. The Fourier transform of the function can be written in terms of the Green's function so that $\frac{d\sigma^V}{dE} \propto \mathrm{Im}\,G_V^\sigma(E)$ with

$$EG_V^\sigma(E) = \frac{1}{2\pi}<[d_{\tau_p\sigma}, d_{\tau_p\sigma}^+]> + <<[d_{\tau_p\sigma}, H]; d_{\tau_p\sigma}^+>>_E$$

where $H$ is the Hubbard Hamiltonian [14]. Thus, the theoretical Auger spectrum in the third Hubbard approximation using the unbounded Lorentz density of states can be determined by the well-known result

$$\frac{d\sigma^V}{dE} \propto \frac{\Delta}{\pi}\left(\frac{1-n_{-\sigma}}{(E-T^V)^2+\Delta^2} + \frac{n_{-\sigma}}{(E-T^V-U)^2+\Delta^2}\right)$$

with the condition $E_c = E_{c'} = 0$, where $n_{-\sigma}$ is the number of electrons with spin $-\sigma$, $T^V$ and $T^V + U$ are the location parameters of the Auger electron peaks, $U$ is the intra-atomic Coulomb correlation energy between electrons of opposite spin, $\Delta$ is the half-bandwidth at the half-maximum peak and $\delta(\cdot)$ is the Dirac delta function [14,15]. Moreover, the imaginary part of the self-energy, i.e.,

$$\mathrm{Im}\,\Sigma^\sigma(E) = \frac{-Un_{-\sigma}(1-n_{-\sigma})\Delta}{[E-T_o-U(1-n_{-\sigma})]^2+\Delta^2}$$

defines the lifetime of the quasi-electrons and shows an additional peak in the Auger spectrum for nearly half-filled bands. Often, the core-valence-valence ($CVV$) Auger line shape can be successfully interpreted as the self-convolution of the valence band of density [16]. Moreover, if the life time of the core hole is short enough, the Auger electron can interact with the particles and fields that are present during the creation of the core hole. This means that different physical effects will change the spectral intensities. Up to this point, the treatment is essentially the same as in the usual approach to the quantum theory. But now, we bring in a basically new feature of Bohm's approach, i.e. that the reality includes Auger electrons that follow well-defined trajectories, as well as the wave function. From the point of view of Auger electrons, each of the non-overlapping parts of the wave function describes a separate "Bohm's channel". If an electron is in the channel, its quantum potential is determined by the channel alone and other channels do not contribute.

Bohm theory explains why an actual Auger transition occurs in a time that is much shorter than the mean lifetime of the quantum state [13]. We can understand why the Auger transition cannot give information about the layer structure. Thus, in our abstruse system, the Auger effect is described by a part of the wave function and is completed by the specification of the $\tau_l$-atomic layer position. Therefore, the mathematical formalism of the modified Bohm's theory is presented with explicit applications to the profiles in the desorption and the adsorption processes.

## 3. Modified Bohm's theory for abstruse measurements

Thus, we assume the following approximations by using the standard method of separation of variables and rescaling depth [nm] to the number of monolayers. Hence, the Hamilton-Jacobi function is given by

$$S_i = -E't + \frac{1}{A}S_i(z) \quad V_i = V_i(z) \quad \text{and} \quad Q_i = Q_i(z)$$

where $z$ is the monolayer depth of the thin film along the $z$-axis and $E'$ and $A$ are constants. Therefore, Equation (1) can be rewritten as

$$V_i(z) + Q_i(z) + \frac{1}{2m_i A}[\nabla S_i(z)]^2 = E' \qquad (3)$$

In order to discuss the $i^{th}$-layer, we introduce the concept of the sum due to the removal of layers due to ion bombardment as well as the adsorbing layers due to the adsorption process. If

$$\sum_i S_i(z) = N_l(z), \quad \sum_i V_i(z) = V_l(z),$$
$$\sum_i C^i Q_i(z) = Q_l(z) \quad \text{and} \quad M = \sum_i \frac{1}{2m_i}$$

where $C^i$ is a constant, and then we finally obtain the modified equation

$$V_l(z) + Q_l(z) + \frac{M}{A}[\nabla N_l(z)]^2 = 2\widetilde{E} \qquad (4)$$

where $N_l(z)$ is the function of the depth profile in a multilayer system and $\widetilde{E}$ is a new constant. In our considerations the asymmetric shape of the Auger electron line is described by the two Dirac delta functions that can be represented by Lorenz functions. This procedure allows us to combine the Auger effect with the quantum potential representation. The quantum potential is given by $Q_l(z) = -\sum_{i=1} C^i \delta(z - i\tau_l)$, where $\delta(\cdot)$ is the Dirac delta function, $i = 1, 2, 3...$ and $\tau_l$ is the source position of the Auger electron emission from the $i^{th}$-layer atoms. This means that the potential is due to the Auger electron effect [2,10]. Moreover, it also assumes that $M = 1$, $V_l(z) = 2V_o$, where $V_o$ is a constant, thus Equation (4) can be expressed by

$$\frac{1}{A}[\nabla N_l(z)]^2 = \widetilde{E} - 2V_o + \sum_{i=1} C^i \delta(z - i\tau_p) \quad (5)$$

The application of the Laplace transform to both sides of Equation (5) yields the solution for the transformed problem as

$$\frac{1}{A}s^2 N_l(s) = \alpha s + \frac{\widetilde{E}}{s} - \frac{2V_o}{s} - Ce^{-\tau_p s} + \\ + C^2 e^{-2\tau_p s} - C^3 e^{-3\tau_p s} + .... \quad (6)$$

with

$$L\{N_l(z)\} = N_l(s), \ N_l(0) = \alpha, \ \frac{dN_l(z)}{dz} = \frac{da}{dz} = 0 \quad (7a)$$

$$L\{\frac{dN_l(z)}{dz}\} = sL\{N_l(z)\} - N_l(0) = s \cdot N_l(s) - a, \ s \in \Re \quad (7b)$$

$$s^2 L\{N_l(z-b)G(z-b)\} = \begin{cases} s^2 e^{-b \cdot s} N_l(s) - as, & b > 0 \\ s^2 N_l(s) - as, & b = 0 \end{cases} \quad (7c)$$

where $L\{N_l(z)\}$ and $N_l(s)$ denote the Laplace transform of the function profile $N_l(z)$, $s$ is the Laplace variable, $\alpha$ and $b$ are constants and $\Re$ is the field of real numbers (for details see [3]). Rearranged we get

$$N_l(s) = A \begin{pmatrix} \frac{\alpha}{s} + \widetilde{E}\frac{2}{s^3} - V_o \frac{2}{s^3} - \frac{C}{s^2} e^{-\tau_l s} + \\ + \frac{C^2}{s^2} e^{-2\tau_l s} - \frac{C^3}{s^2} e^{-3\tau_l s} + ..... \end{pmatrix} \quad (8)$$

The value parameters for the solution of the transformed problem can be determined by the experimental data. The joint inverse Laplace transform of the (8), i.e. $L^{-1}\{N_l(s)\} = N_l(z)$, gives

$$N_l(z) = A(\alpha H(z) + \widetilde{E}z^2 - V_o z^2 - C(z - \tau_l)H(z - \tau_l) + \\ + C^2(z - 2\tau_l)H(z - 2\tau_l) - ....) \quad (9)$$

where $H(\cdot)$ is the Heaviside step function. This solution of the discontinuous function satisfies Equation (5) and will enable us to calculate the probability density (Eq. (2)). Thus, if $P_l = P_l(z)$, $P_l(z) = \sum_i R_i^2 = \sum_i |\psi_i|^2$ and $div(-P_l(z)\nabla N(z)2M) = 0$, then the probability density function can be expressed as

$$P_l(z) = -\frac{const}{2M\nabla N_l(z)} \Rightarrow \nabla N_l(z) < 0 \quad (10)$$

where $N_l(z)$ is defined by Equation (9). This means that the probability density function predicts the multilayer structure. The negative critical value of $P_l(z)$ predicts the kind of break or „cut" between the physical and unphysical layer systems. We see that Eqs. (5) and (10) are the main theoretical results and show the modified formalism for the solution of abstruse measurements. Of course, different forms of the potential provide different results. We attempt to apply to other potential forms, for instance $Q_l(z) = -\sum_{i=1} C^i H(z - i\tau_l)$, $Q_l(z) = -\sum_{i=1} C_i H(z - i\tau_l)$, $Q_l(z) = -\sum_{i=1} C_i \delta(z - i\tau_l)$ etc. It is easy to check whether the step-like function

$$N_l(z) = A(\alpha H(z) + \widetilde{E}z^2 - V_o z^2 - \frac{C}{2}(z - \tau_l)^2 H(z - \tau_l) + ....)$$

satisfies the differential equation

$$\frac{1}{A}[\nabla N_l(z)]^2 = \widetilde{E} - 2V_o + \sum_{i=1} C^i H(z - i\tau_l) \ .$$

Note that the last term is due to the quantum potential and the $C^i$ – factor is the amplitude. A natural generalisation of the monolayer depth profile problem is $N(z) := \sum_l c_l N_l(z)$, where $c_l$ is the molar fraction of the $l$ – element. It is worth pointing out that the present formalism indeed leads to the solution of the abstruse measurements. We now apply this formalism to two classes of experiments. We will show that the abstruse layer profiles can be converted to real layer structures.

## 4. Theory and experiments in abstruse measurements

The Auger electron spectroscopy system consists of an ultrahigh vacuum system, a primary electron gun for specimen excitation, an ion gun, an energy analyser to detect the Auger electron peaks and a computer for data storage [1,2]. The measurement was carried out in an SP-2000 1/M type vacuum system on a stationary Auger SEA 02 spectrometer [9-12]. The sample was excited by bombardment with the primary electron energy $E_p = 3$ keV and the current $I_p = 3$ μA from the electron gun. These electrons cause (among other interactions) the emission of Auger electrons with characteristic energies. The Cylinder Mirror Analyser (CMA) consists of two coaxial cylinders with the inner cylinder at ground potential and a potential of $-V_{exp}$ on the outer cylinder. The electrons that are emitted from the source of the Auger electrons on the axis move in the field-free space towards a mesh-covered annular aperture in the inner cylinder, while those within the angular spread and pass into the space between the cylinders. These electrons, which have kinetic energy, are deflected by the potential of the outer cylinder through a second mesh-covered aperture and are focused at the collector current as well as on the axis. A general consideration of the current $I(V_{exp} + \Delta V_{exp})$, where $\Delta V_{exp} = V_{const} \sin(\omega t)$ is the perturbing voltage, can be rewritten as a Taylor series, $I(V_{exp} + \Delta V_{exp}) \sim I_o(V_{exp}) + I'(V_{exp})[V_{const}\sin(\omega t)] - I''(V_{exp})[V_{const}\cos(2\omega t)] + ...$. The term $I_o(V)$ corresponds to the part that does not vary with time. Next, the fundamental frequency term is proportional to $I'(V)$, i.e. to the energy distribution $N(E)$ (for example, $\mathrm{Im}\, G_V^\sigma(E) \cong N_V(E)$, Auger valence-band spectrum). The second harmonic term $(2\omega)$ is $-I''(V)$ gives the derivative of the energy distribution $dN(E)/dE$. Simply, scanning the potential $-V_{exp}$ on the outer cylinder of the CMA directly gives the energy distribution of the electrons pass through it, because the transmission of the CMA varies as $E$, the recorded distribution is not $N(E)$, but $EN(E)$ [1]. Similarly, the differential distribution will be not be just $dN(E)/dE$, but $EdN(E)/dE$. Thus, for a reasonable small modulating voltage ($V_{const} = 1 V_{p-p}$), the second harmonic term ($f = 5$ kHz) in the collector current enables us to plot the derivative $EdN(E)/dE$ (see Fig. 1). Because of the small Auger signals, studies are usually carried out in the derivative mode. The intensity of the Auger peak can be measured as the peak-to-peak-height (APPH) in the differential spectrum. We have already pointed out that even if a monoenergetic beam of the Auger electrons is injected into the analyzer, full-width at half-maximum Auger peak is recorded due to various instrumental limitations. The analyser transmission and resolution were 8% and the analyser about 0.6%, respectively. The energy scale was calibrated by measuring the analyser voltage that is required to transmit the elastically reflected primary electrons of a known energy. In practice, all of the elements, with the exception of hydrogen and helium, produce Auger electrons; detection sensitivity is typically between 1 and 0.1 percent of a monolayer.

The carbon/passive layer on the NiTi substrate system was studied by "in situ" removing the carbon atoms using argon ion spattering and carefully selecting the sputter parameters. The investigated NiTi alloy was produced by Krupp (Germany) with a composition of Ti 51.1 at.% Ni 48.9 at.% and the parent phase structure [17,18]. The passive layer was achieved after heating at 500°C for five minutes in the air atmosphere. The carbon was deposited by spontaneous adsorption from the air at room temperature [1,9]. This carbon film was studied.

By monitoring the APPH signal from the desorbed elements as a function of the sputter time $t_{sputter}$, we obtain the experimental data with an estimated error limit of 7% (Fig. 2, top panel). Conversion of the sputter time into the monolayer depth in the sputter depth profile can be determined by the relationship $s \approx v_{sputter} t_{sputter}$, where $v_{sputter}$ is the sputter velocity (Fig. 2, middle panel) [1,9]. We see that the sputter depth profile analysis shows the abstruse measurements (Fig. 2, mid panel). This measurement problem can be solved using the modified Bohm's theory. Thus, Equation (8) was fitted to the experimental data (Fig. 2, middle panel). Consequently, we obtain the solution of the transformed problem

$$N_C(s) \cong 0.87 \begin{pmatrix} \dfrac{1.6}{s} - 0.08\dfrac{2}{s^3} - \dfrac{1.25}{s^2}e^{-s} + \\ + \dfrac{1.25^2}{s^2}e^{-2s} - ... - \dfrac{1.25^9}{s^2}e^{-9s} \end{pmatrix} \quad (11)$$

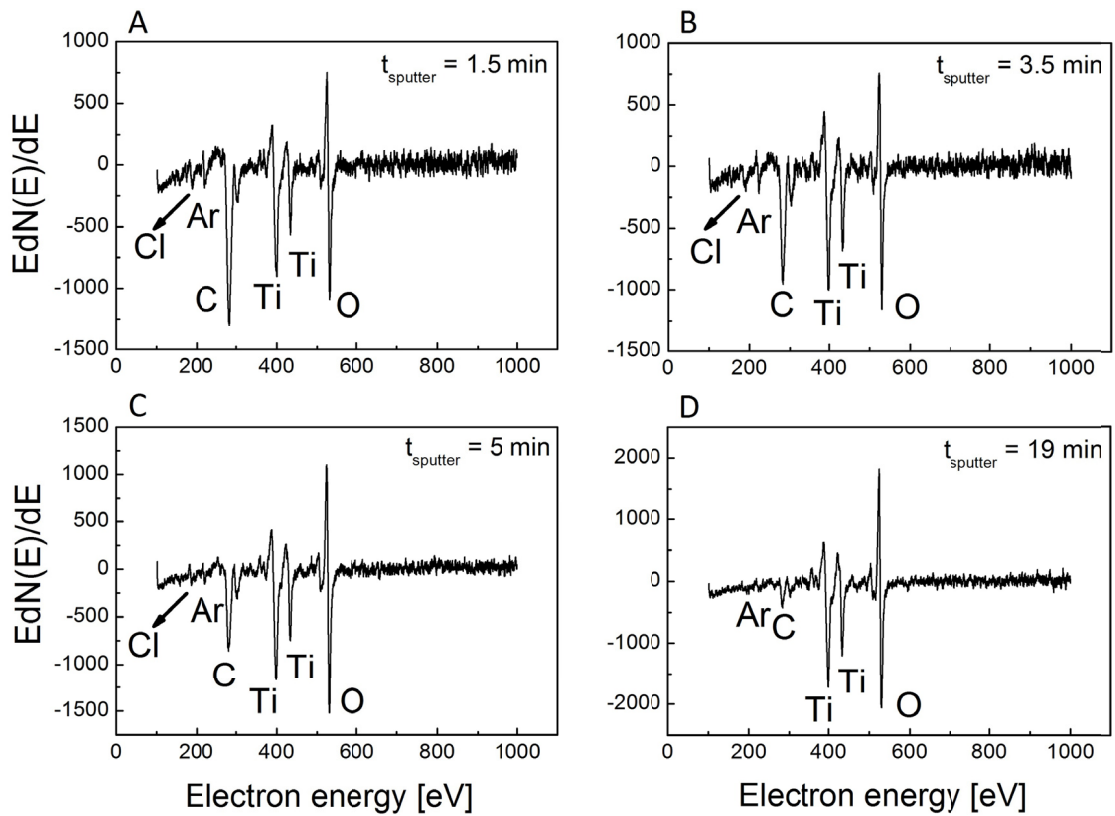

Fig. 1. Experimental Auger spectra for the desorbed carbon/ passive layer system. The values of the sputter time are presented in the upper right-hand corner of each panel

Substituting the fitted parameters into Equation (9), we have

$$N_C(z) \cong \begin{cases} 0.87(1.6 - 0.08z^2) & 0 \le z < 1 \\ 0.87[1.6 - 0.08z^2 - 1.25(z-1)] & 1 \le z < 2 \\ 0.87[1.6 - 0.08z^2 - 1.25(z-1) + 1.25^2(z-2)] & 2 \le z < 3 \end{cases} \quad (12)$$

where $z$ represents the monolayer depth with an estimated error limit of 4% (in the carbon monolayer).

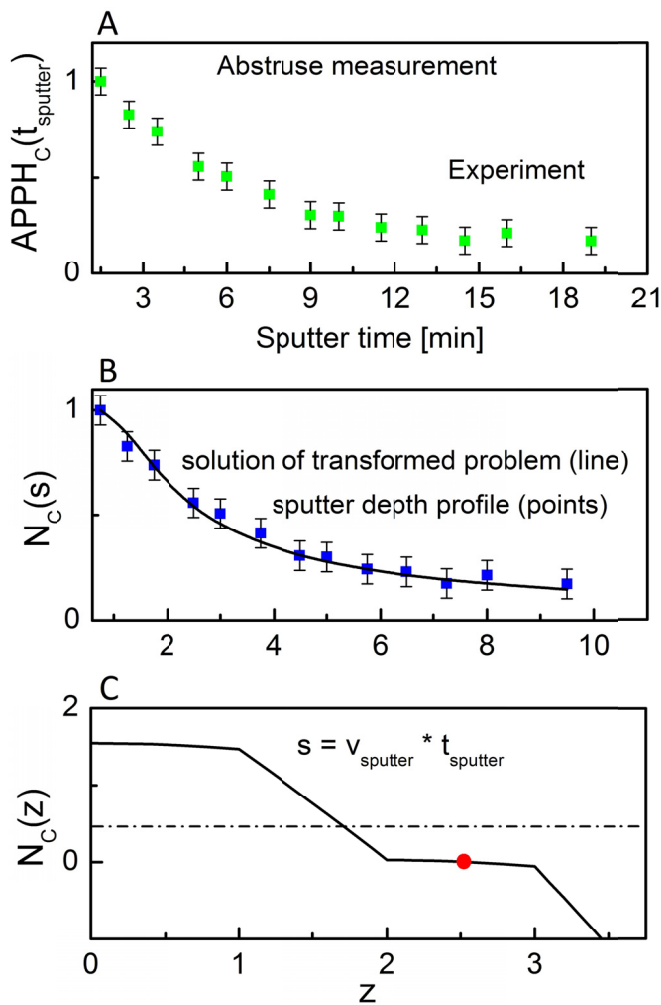

Fig. 2. Results for the desorbed carbon/passive layer on an NiTi substrate. The experimental data (points) are illustrated in A. Conversion of the sputter time into depth ($s \approx v_{sputter} t_{sputter}$, where $v_{sputter} \cong 0.5$ carbon monolayer/min) in the sputter depth profile (points) is plotted in B. The fitted function (solid line) gives the following parameter values: $A \cong 0.87$, $\tilde{E} = 0.01$ eV, $V_o = 0.09$ eV, $M = 1$ kg$^{-1}$, C = 1.25, $\tau_C = 1$, $\alpha = 1.6$ and $i = 1,2,3,...,9$. Parameter $A$ is a dimensionless constant, $\tilde{E}$ is the total energy of Auger electron, $V_o$ is the potential energy of this electron, $M$ is the sum of the inverse mass of particles (electrons), $C$ and $\alpha$ is the dimensionless constant and $\tau_C$ is the source position of the Auger electron emission from the carbon layer. The interpolation of Equation (8) was used to receive Equation (11) with the help of nonlinear fit as given in Microcal Origin. The solution function of Equation (5) is shown in panel C (see Equation (12)). The large dots denote the critical value of the solution between the positive and negative signs

Equation (12) shows the solution of Equation (5), Figure 2C. The probability density function ($P_C(z)/const = -1/[2\nabla N_C(z)]$), where $N_C(z)$ is defined by the Equation (12)) and the inverse probability density function ($const/P_C(z) = -2\nabla N_C(z)$) shows the typical layer structure and denotes the real number of a multilayer system (Fig. 3). Moreover, we can also observe a kind of break or "cut" between the physical and unphysical layers. Therefore, the carbon film has a thickness of four layers.

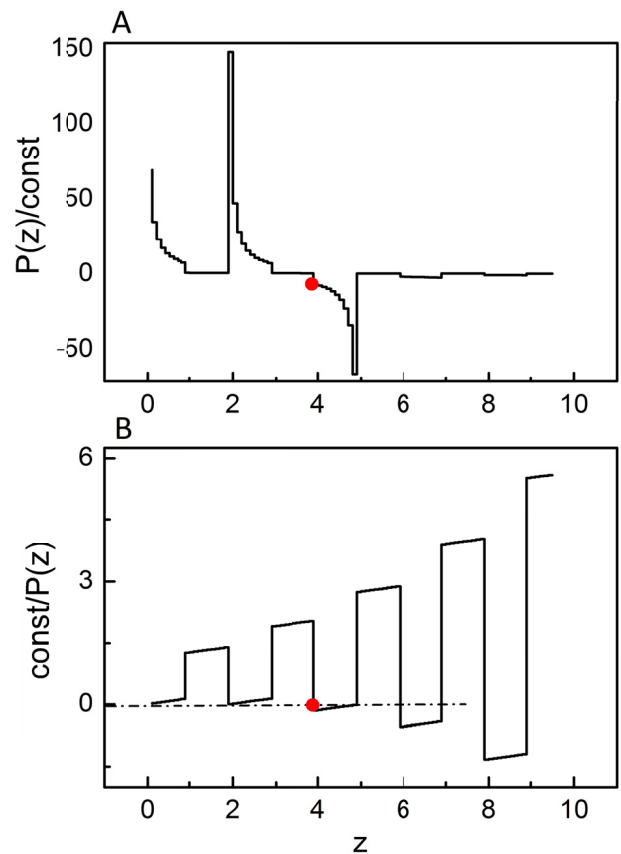

Fig. 3. Probability density function (A) and the inverse probability density function (B) for the same physical system as in Figure 2 using Equations (10) and (12). The large dots denote the critical values between the physical and unphysical layers

The second experimental system is based on the concept of adsorbed oxygen on sites on the smooth surface (roughness was of 8.7 nanometer) of the TiNi alloy so that the coverage varies from zero to one monolayer with an estimated error limit of 5% (Fig. 4). The monolayer is found when the Auger signal is saturated. These measurements were also carried out with the Auger spectrometer using the same material. The mathematical formalism describes the system for monolayer coverage profile, i.e., $i = 1 \Rightarrow S_1(z) = N_{Ti}(z)$ (see Equation (5) and Fig. 4). The results show that the main oxidation processes (Ti → TiO → TiO$_2$) can be separated using the present formalism (for details see [10]). This means that the formalism can also be applied to the abstruse phase transition due to the adsorption (desorption) process. From the theoretical point of view, the quantum potential plays a key role in the solution of abstruse measurements. In summary, the present formalism enables us to obtain the desired results. Of course, the results can also be verified easily by other well-known studies [1,2,5-12,19,20].

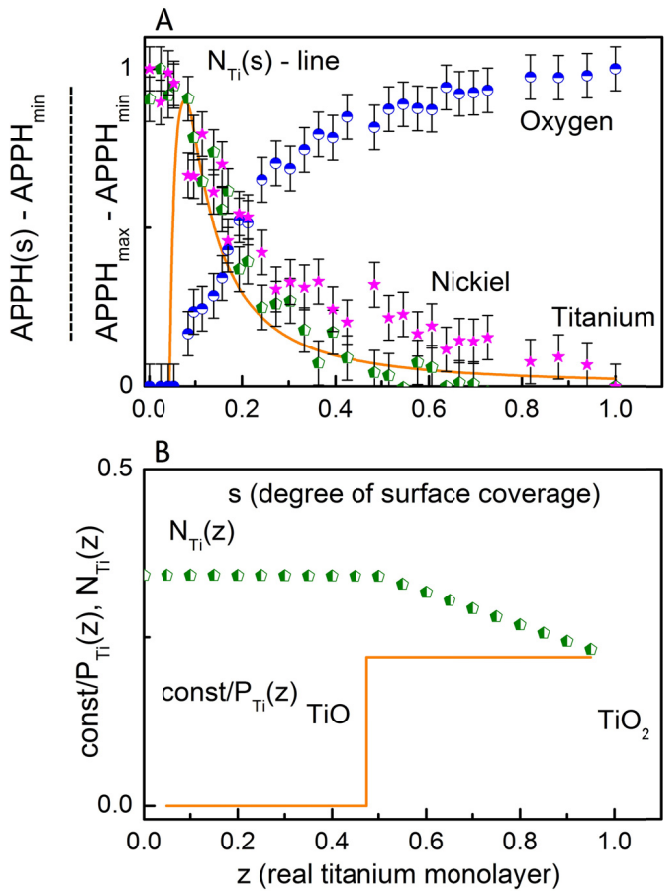

Fig. 4. Results for the adsorbed oxygen/surface of the NiTi alloy in the oxygen medium at a pressure of $10^{-9}$ hPa. The experimental data (points) are illustrated in panel A. The fitted function (line) for the normalised intensity of titanium gives the following parameter values: $A \cong 0.056$, $\widetilde{E} = 0.001$ eV, $V_o = 0.006$ eV, $M = 1$ kg$^{-1}$, $C = 0.2$, $\tau_{Ti} = 0.5$, $\alpha = 0.23$ and $i = 1$ (see Equation (11)). The solution for one monolayer of Equation (5) and the inverse probability function are shown in panel B. The discontinuity shows the critical value between the oxidation processes

## 5. Conclusion

Our final conclusion is that the present formalism of the modified Bohm's theory can be applied to solve abstruse measurements, i.e. the abstruse-multilayer depth-profile can easily be converted to a real layer structure; the abstruse-monolayer coverage profile can be converted to different oxidation processes. Such a parameter theory could lead to new insights and experiments. In principle, the present formalism can be applied as a powerful tool for analysis in other systems.